\documentclass[pra,reprint,aps,twocolumn,showpacs,superscriptaddress,floatfix,amsmath,amssymb,nofootinbib]{revtex4-2}
\usepackage{amsmath,amssymb,graphicx,braket}
\usepackage{hyperref}
\usepackage{xcolor}

\newcommand{\beq}{\begin{equation}}
\newcommand{\eeq}{\end{equation}}
\newcommand{\beqa}{\begin{eqnarray}}
\newcommand{\eeqa}{\end{eqnarray}}

\newcommand{\appropto}{\mathrel{\vcenter{
  \offinterlineskip\halign{\hfil$##$\cr
    \propto\cr\noalign{\kern2pt}\sim\cr\noalign{\kern-2pt}}}}}
    
\begin{document}

\title{Quantum control of tunable-coupling transmons using dynamical invariants of motion}

\author{H. Espin\'os}
\affiliation{Departamento de F\'isica, Universidad Carlos III de Madrid, Avda. de la Universidad 30, 28911 Legan\'es, Spain}
\author{I. Panadero}
\affiliation{Departamento de F\'isica, Universidad Carlos III de Madrid, Avda. de la Universidad 30, 28911 Legan\'es, Spain}
\affiliation{Arquimea Research Center, Camino las Mantecas s/n, 38320 Santa Cruz de Tenerife, Spain}
\author{J. J. Garc\'ia-Ripoll}
\affiliation{Instituto de Física Fundamental (IFF), CSIC, Calle Serrano 113b, 28006 Madrid, Spain}
\author{E. Torrontegui}
\affiliation{Departamento de F\'isica, Universidad Carlos III de Madrid, Avda. de la Universidad 30, 28911 Legan\'es, Spain}
\email{eriktorrontegui@gmail.com}

\begin{abstract}
We analyse the implementation of a fast nonadiabatic CZ gate between two transmon qubits with tuneable coupling. The gate control method is based on a theory of dynamical invariants which leads to reduced leakage and robustness against decoherence. The gate is based on a description of the resonance between the $\ket{11}$ and $\ket{20}$ using an effective Hamiltonian with the 6 lowest energy states. A modification of the invariants method allows us to take into account the higher-order perturbative corrections of this effective model. This enables a gate fidelity several orders of magnitude higher than other quasiadiabatic protocols, with gate times that approach the theoretical limit.
\end{abstract}

\maketitle

\section{Introduction}\label{Introduction}

Superconducting qubits are currently one of the most promising platforms to perform highly scalable quantum computations. In the last two decades, there has been great progress both in the quantity of qubits and the quality of their operations~\cite{Wilhelm2008,Barends2014,Gambetta2016a,DiCarlo2017}. This progress has been largely made possible thanks to the transmon qubit~\cite{Koch2007}, a relatively simple, easy to reproduce design, with competitive decoherence times~\cite{Nersisyan2019,Place2021}. Recent progress in quantum gates with transmons have led to substantial speed-ups, surpassing coherence times by up to three orders of magnitude, with competitive fidelities that approach the requirements for scalable quantum error correction~\cite{Nissim2016,Gambetta2016,Hu2019} and fault-tolerant quantum computation~\cite{Chow2015,Martinis2015,Vuillot2017}. 
The implementation of two-qubit gates with superconducting qubits is possible with many different strategies: gates assisted by microwave controls~\cite{Steffen2011}, parametrically modulated qubits~\cite{Reagor2018} and couplers~\cite{McKay2016,Ganzhorn2019}, gates implemented with tunable-frequency qubit-qubit resonances~\cite{Rol2019} or tunable couplings~\cite{Chen2014,Mundada2019,Han2020,Xu2020,Collodo2020,Ye2021}.
These demonstrations have proved successful at performing CZ~\cite{Barends2014,Chen2014,Rol2019,Xu2020,Ye2021}, CPHASE~\cite{Collodo2020} and iSWAP~\cite{Han2020} gates with fidelities over 99\%.

Using a tuneable coupling to implement quantum gates presents the advantage of being more energy efficient, requiring less control lines, and having better coherence in general~\cite{Chen2014}. The use of tuneable couplers allows isolating the qubits and cancelling parasitic interactions during single-qubit gates and rest periods. Tuneable couplings also eliminate the problem of frequency crowding that arises in other two-qubit gate strategies with always-on interactions~\cite{Mundada2019}. In order to manipulate the qubits, most of the current protocols benefit from adiabatic controls~\cite{Barends2014,Chen2014,Mundada2019,Collodo2020,Xu2020}. Although adiabatic processes are robust against control imperfections, they are intrinsically slow, limiting the number of operations that can be performed within the lifespan of the qubits, and thus leaving the system more vulnerable to incoherent errors. 
This has motivated the development of nonadiabatic protocols~\cite{Torrontegui2013,Guery-Odelin2019} using the frequency-tunable, fixed-coupling architecture~\cite{Barends2019,Li2019,Garcia-Ripoll2020}. The principle of these nonadiabatic designs is to allow transitions along the process that implements the gate, but suppress them once said process is finished. 
A common technique to achieve this is to drive the computational states using dynamical invariants of motion~\cite{Lewis1969}, such that the system's Hamiltonian and the invariant share eigenstates at the beginning and at the end of the control~\cite{Chen2010}.

In this work, we present a new method to implement universal CZ gates between transmon connected by a tunable coupler. The formalisim is based on an exact technique that uses invariants of motion to inverse-engineer the gate protocol, and is made possible by a simpler formulation of the transmon dynamics in a reduced space. The derivation of this effective Hamiltonian, starting from the standard two-transmon Hamiltonian, is carried out in Sect. \ref{model}. Using this simplified version, in Sect. \ref{CZgate} we design faster-than-adiabatic protocols to perform the CZ gate. We compare two different approaches to engineer the turn-on and off processes of the coupling: fast quasiadiabatic dynamics (FAQUAD) and invariant-based inverse engineering. In Sect. \ref{test}, we  numerically simulate the two-transmon system to evaluate the drivings performance and identify possible sources of infidelity. With the aid of the effective description we introduce higher order energy corrections to rectify incoherent errors due to an energy Stark-shift, showing that the invariant-based method has a great resilience, bringing the gate infidelity below $10^{-5}$ in competitive operation times. Finally, Sect. \ref{conclusions} contains the manuscript main conclusions.

\section{Transmon model}\label{model}
\subsection{Single transmon}\label{single}
A transmon is a charge-based superconducting qubit that is created by shunting a Josephson Junction with a large capacitor. This leads to a drastic reduction in sensitivity to charge in the superconducting island, while keeping an anharmonicity that separates the qubit subspace from the rest of energy levels~\cite{Koch2007}. The Hamiltonian describing one ``bare" transmon, in the number-phase representation, reads
\beq
  \label{H0}
  \hat H_T = 4E_C \hat n^2 - E_J \cos(\hat \varphi), 
\eeq
with $\hat n$ and $\hat \varphi$ satisfying the canonical commutation relation $[\hat n,\exp(i\hat \varphi)]=-\exp(i\hat\varphi)$. $E_C$ is the charging energy, related to the total capacitance $C$ of the transmon qubit as $E_C=2e^2/C$, and $E_J$ is the Josephson energy. A transmon typically operates at a large $E_J/E_C$ ratio ($\gtrsim 50$), where the qubit dynamics are analogous to a massive particle in a weakly anharmonic potential, 
\beq
\label{anharmonic_oscillator}
  \hat H_T = 4E_C \hat n^2 + \frac{E_J}{2} \hat \varphi^2-\frac{E_J}{4!}\hat \varphi^4 +\cdots . 
\eeq
Identifying $\hat p^2/2m \sim 4E_C\hat n^2$ and $m\omega^2 \hat x^2 \sim E_J\hat \varphi^2$, the quadratic part of the Hamiltonian can be diagonalized using the ladder operators of the harmonic oscillator, $\hat a$ and $\hat a^\dagger$, defined as,
\beq
\label{a_adagger}
\hat n = i\left(\frac{E_J}{8E_C}\right)^{1/4}\frac{(\hat a-\hat a^\dagger)}{\sqrt{2}}, \hspace{0.4cm}
\hat \varphi = \left(\frac{8E_C}{E_J}\right)^{1/4}\frac{(\hat a+\hat a^\dagger)}{\sqrt{2}}.
\eeq
Notice how, at large $E_J/E_C$, $\hat \varphi$ decreases in magnitude, and hence larger powers of $\hat \varphi$ can be neglected in the expansion from Eq.\ \eqref{anharmonic_oscillator}. Introducing Eq.\ \eqref{a_adagger} into this expansion, 
the transmon Hamiltonian reads
\begin{eqnarray}
 \label{H0_HO}
 \hat H_T \simeq \omega_{01}\hat a^\dagger \hat a&+&\frac{\alpha}{2}\hat a^{\dagger 2}\hat a^2+\frac{\alpha}{12}\left(\hat a^4+\hat a^{\dagger 4}\right. \nonumber\\
 &+& \left. 4\hat a^{\dagger 3 }\hat a + 4\hat a^\dagger \hat a^3 +6\hat a^{\dagger 2}+6\hat a^2\right),
\end{eqnarray}
where $\omega_{01}\simeq\sqrt{8E_{C}E_{J}}-E_J$ represents the 
splitting between the two lowest energy states, $\ket{0}$ and $\ket{1}$. The anharmonicity $\alpha\simeq-E_{C}$ is small but allows us to detune all higher energy states, $\ket{2},\ket{3},\ket{4},\ldots$ from the qubit subspace. 

The eigenstates of the transmon Hamiltonian, denoted by $\ket{n}$, differ from the eigenstates of the harmonic oscillator due to all the anharmonic terms in Eq.\ \eqref{H0_HO} that do not preserve the number of excitations, i.e., those with different number of $\hat a$ and $\hat a^\dagger$ operators. In fact, these counter-rotating terms can be treated as a perturbation to the reference Hamiltonian $\hat H_0=\omega_{01}\hat a^\dagger \hat a+(\alpha/2)\hat a^{\dagger 2}\hat a^2$, whose eigenstates are those of the harmonic oscillator. Using standard time-independent perturbation theory at first order in $\alpha/\omega_{01}$, we find for the lowest-energy states
\begin{align}
\label{perturbative_states}
\ket{0} \simeq &\ket{\Psi_0}- \frac{\sqrt{2}\alpha}{3(2\omega_{01}+\alpha)}\ket{\Psi_2}-\frac{\sqrt{6}\alpha}{12(2\omega_{01}+3\alpha)}\ket{\Psi_4},\nonumber\\
\ket{1} \simeq &\ket{\Psi_1}- \frac{5\sqrt{6}\alpha}{6(2\omega_{01}+3\alpha)}\ket{\Psi_3}-\frac{\sqrt{30}\alpha}{12(2\omega_{01}+5\alpha)}\ket{\Psi_5},\nonumber\\
\ket{2} \simeq &\ket{\Psi_2} +\frac{\sqrt{2}\alpha}{3(2\omega_{01}+\alpha)}\ket{\Psi_0}- \frac{8\sqrt{3}\alpha}{3(2\omega_{01}+5\alpha)}\ket{\Psi_4}\nonumber\\
&-\frac{\sqrt{10}\alpha}{2(4\omega_{01}+15\alpha)}\ket{\Psi_6},
\end{align}
where $\ket{\Psi_i}$ represents the $i$-th eigenstate of the harmonic oscillator. The perturbative expansion reveals that, when an interaction term proportional to $\hat n$ is present, both $\hat a$ and $\hat a^\dagger$ produce a coupling between two adjacent states, in opposition to what happens in the harmonic oscillator. For instance, the operator $\hat a$ projects the state $\ket{0}$ into state $\ket{1}$ with an amplitude proportional to $\alpha/\omega_{01}$.  

\subsection{Coupled transmons}\label{coupled}
In this work we consider two transmons with a tunable interaction between them using the design demonstrated in~\cite{Chen2014}. In this design, two superconducting Xmon qubits are coupled through a circuit that uses a single flux-biased Josephson junction and acts as a tunable current divider. The physics behind this tunable coupler is well explained using a simple linear model (see Ref.~\cite{Geller2015} for a full discussion),
\beq
  \label{H}
  \hat H(t)= \hat H_{T,a}+\hat H_{T,b} +g_C(t)\hat n_a\hat n_b,
\eeq
where $g_C$ is the tunable coupling that embodies an interaction that can be varied continuously with nanosecond resolution from negative to positive, going smoothly through zero, where the transmons are isolated. $\hat H_{T,i}$ is the Hamiltonian given by Eq.\ \eqref{H0} of each ``bare" transmon $i=a,b$ used to encode a qubit. 

For typical experimental circuit parameters, the coupling takes values in the MHz range, while $\omega_{01}$ is usually in the GHz range. Due to this energy mismatch, only close-to-degeneracy states will be able to interact and produce transitions from one state to another. Assuming that both transmons have similar frequencies, these close-to-degeneracy states are the ones that share the total number of excitations, for instance, states $\ket{01}$ and $\ket{10}$, or states $\ket{11}$, $\ket{02}$ and $\ket{20}$. In particular, denoting the ladder operators as $\hat a$, $\hat a^\dagger$ for the first qubit, and $\hat b$, $b^\dagger$ for the second qubit, we can compute the non-zero matrix elements of the coupling between these states using the perturbative expansions from Eq.\ \eqref{perturbative_states}, 
\begin{align}
\widetilde{J}_1(t)&\equiv\bra{01}g_C(t)\hat n_a \hat n_b\ket{10} \nonumber\\
&= J(t)\bra{01}(-\hat a^\dagger \hat b^\dagger + \hat a^\dagger \hat b + \hat a \hat b^\dagger -\hat a\hat b )\ket{10}\nonumber\\
&= J(t)\left[1+\frac{2\alpha_a}{3(2\omega_a+\alpha_a)}+\frac{2\alpha_b}{3(2\omega_b+\alpha_b)}\right.\nonumber\\
&\left.+ \mathcal{O}\left(\frac{\alpha_{a,b}^2}{\omega_{a,b}^2}\right) \right],
\end{align}
\begin{align}
\widetilde{J}_2(t)&\equiv \bra{11}g_C(t)\hat n_a \hat n_b\ket{02} \nonumber\\
&= \sqrt{2}J(t)\left[1+\frac{-\alpha_b}{3(2\omega_b+\alpha_b)}+\frac{5\alpha_b}{2(2\omega_b +3\alpha_b)}\right.\nonumber\\
&\left.+\frac{2\alpha_a}{3(2\omega_a+\alpha_a)}+ \mathcal{O}\left(\frac{\alpha_{a,b}^2}{\omega_{a,b}^2}\right) \right],
\end{align}
\begin{align}
\widetilde{J}_3(t)&\equiv\bra{11}g_C(t)\hat n_a \hat n_b\ket{20} \nonumber\\
&= \sqrt{2}J(t)\left[1+\frac{-\alpha_a}{3(2\omega_a+\alpha_a)}+\frac{5\alpha_a}{2(2\omega_a +3\alpha_a)}\right.\nonumber\\
&\left. +\frac{2\alpha_b}{3(2\omega_b+\alpha_b)} +\mathcal{O}\left(\frac{\alpha_{a,b}^2}{\omega_{a,b}^2}\right) \right],
\end{align}
 where $ J(t)=(1/2)g_C(t)\left[E_{J,a}E_{J,b}/(64 E_{C,a}E_{C,b}\right)]^{1/4}$. For typical experimental values, these first order corrections represent 5-10\% of the total coupling, and they become smaller as the ratios $\alpha_i/\omega_i$ decrease.

As we already mentioned, due to energy mismatch of 2-3 orders of magnitude between the frequency of the transmons $\sim$ GHz and the coupling $\sim$ MHz, states with different number of excitations are well separated in energy and are not affected by the coupling term. For instance, the $\hat a^\dagger\hat b^\dagger$ term coming from Hamiltonian \eqref{H} couples states $\ket{00}$ and $\ket{11}$ with an amplitude $\sim J(t)$. Eliminating this term leads to corrections in the energy of said states of order $J^2(t)/(\omega_a+\omega_b)\ll$ GHz and can be safely neglected. Therefore, in the basis of eigenstates of the uncoupled problem, $\{\ket{00}, \ket{01}, \ket{10}, \ket{02}, \ket{11}, \ket{20}\}$, the Hamiltonian \eqref{H} is very well approximated by a Hamilonian matrix of the form
\beq
  \label{Heff}
  \hat H(t)  =\left( \begin{matrix}
      0 & 0 & 0 & 0 & 0 & 0 \\
      0 & \omega_b & \widetilde{J}_1(t) & 0 & 0 & 0 \\
      0 & \widetilde{J}_1(t) & \omega_a & 0 & 0 & 0 \\
      0 & 0 & 0 & 2 \omega_b+\alpha_b & \widetilde{J}_2(t)& 0 \\
      0 & 0 & 0 & \widetilde{J}_2(t) & \omega_a+\omega_b & \widetilde{J}_3(t) \\
      0 & 0 &0 & 0 & \widetilde{J}_3(t) & 2\omega_a + \alpha_a
    \end{matrix}\right).
\eeq
With the structure of Eq.\ \eqref{Heff}, we emphasised that the three subspaces $S_1:=\{\ket{00} \}$, $S_2:=\{\ket{01}, \ket{10}\}$, and $S_3:= \{ \ket{02}, \ket{11}, \ket{20}\}$, each of them with different number of excitations, are driven independently. In order to perform a certain operation, the coupling $g_C(t)$ can be engineered to produce the desired dynamics in each of these subspaces. Henceforth, we will design $J(t)$ instead of $g_C(t)$ for simplicity (they only differ in a scale factor).

\begin{figure}[t]
    \centering
    \includegraphics[width=0.9\linewidth]{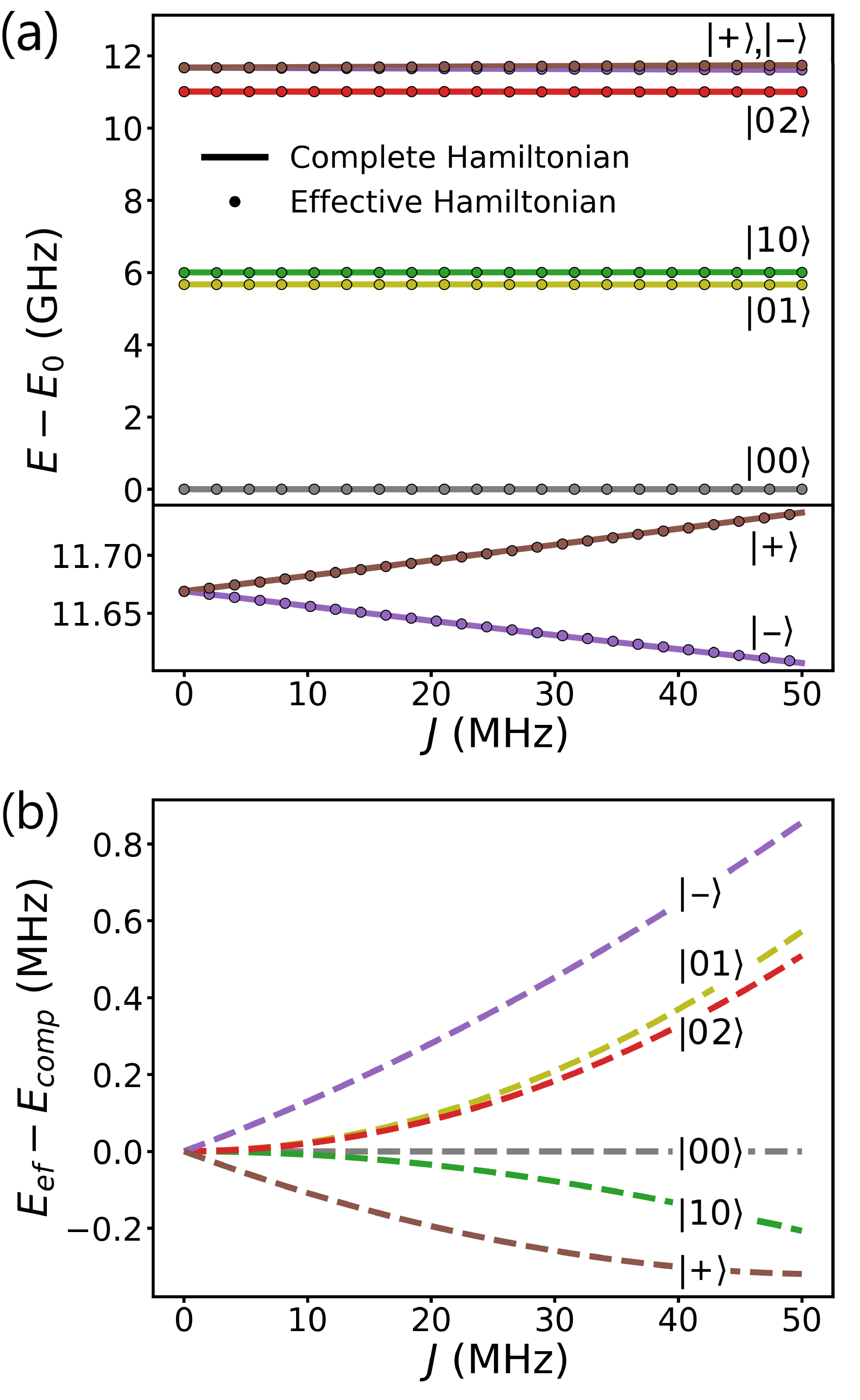}
  \caption{(a) Numerically calculated lowest eigenenergies of Hamiltonians (\ref{H}) (solid lines) and (\ref{Heff}) (round markers) as a function of the coupling strength between the two transmons. The lower panel shows a zoom-in at the avoided crossing of levels $\ket{11}$ and $\ket{20}$, that are degenerate when there is no coupling ($J=0$) and split into two dressed states $|\pm\rangle$ as the coupling strength grows. (b) Energy difference between the effective Hamiltonian and the complete Hamiltonian levels.}
  \label{fig:Energies}
\end{figure}
%
%
%

In Fig.\ \ref{fig:Energies} we test the validity of the approximations made to find the Hamiltonian \eqref{Heff} by comparing its eigenenergies with the lowest six numerically calculated eigenenergies of the complete Hamiltonian \eqref{H} as a function of the coupling strength $J$, not taking into account its time dependence. As shown in Fig.\ \ref{fig:Energies}(b), the effective Hamiltonian reproduces the lower energy levels of the two-transmon system with very little deviation, despite having neglected interactions between two-qubit states with a different number of excitations. In the following section, we will derive the different control protocols from \eqref{Heff} to construct a CZ gate, although the full Hamiltonian \eqref{H} will be simulated with the experimental parameters~\cite{Chen2014}:
\begin{subequations}
\begin{eqnarray}
 \omega_a &=& 2\pi \times 6.00\text{ GHz,}\\
 \omega_b &=& 2\pi \times 5.67\text{ GHz},\\
 \alpha_a&=&\alpha_b =-2\pi\times\text{ 0.33 GHz},\\
 J_M&=&2\pi\times 16.0\text{ MHz},
\end{eqnarray}
\end{subequations}
with $J_M$ the maximum coupling. Thus, small deviations such those shown in Fig.\ \ref{fig:Energies}(b), shall be numerically corrected when designing the protocols to achieve maximum fidelity. 

\section{The {\it CZ} gate}\label{CZgate}
The CZ gate is simple to implement and can readily generate controlled-NOT (CNOT) logic. Acting on the computational basis $\{\ket{00}, \ket{01}, \ket{10}, \ket{11}\}$, this gate generates the transformation $\hat U^{CZ}=\mbox{diag}(1,1,1,-1)$, meaning that it leaves the first 3 states of the computational basis unchanged and adds a $\pi$-phase to the last one.
The CZ gate has been already experimentally demonstrated using transmons qubits where 
the tunable frequency of one transmon controls the dynamics~\cite{Dicarlo2009,Barends2014,Rol2019}, and with transmons connected through a tunable coupler~\cite{Chen2014,Xu2020,Ye2021}.
A recent work~\cite{Garcia-Ripoll2020} showed that, among the different possible
protocols that implement the gate, those inspired on invariants and variational methods lead to more experimentally friendly controls such as better properties of finite bandwidth and resilience to discretization, optimal control of leakage outside the computational basis, and a greater robustness against decoherence. Using quantum control, the performance of this gate can be improved independently of the control design by a proper optimization of the waiting time and destination frequency~\cite{Garcia-Ripoll2020}.
%
%
%
\begin{figure}[t]
  \centering
  \includegraphics[width=0.95\linewidth]{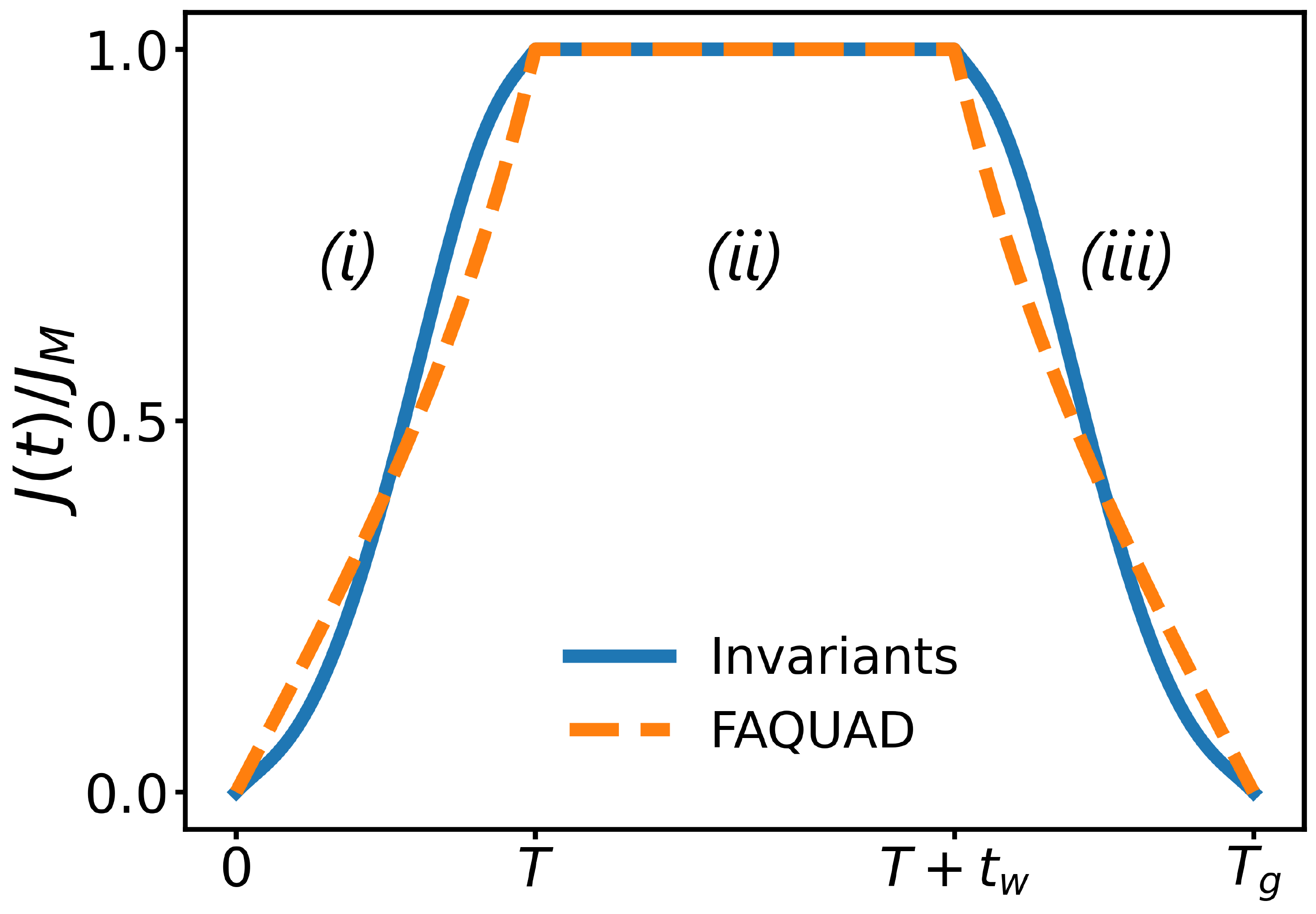}
  \caption{Gate protocols for the CZ gate: two uncoupled transmons are placed at the degeneracy points $\omega_a=\omega_b-\alpha_a$ and the control $J(t)$ is turned on in a time $T$. During a 
  time $t_w$, both transmons interact until the desired gate is implemented. The coupling is symmetrically turned off. The shape of the turn on/off process depends on the design approach. We show sample curves of the FAQUAD and invariant protocols.}
  \label{fig:control}
\end{figure}
%
%
%
\subsection{Design of the gate}
Here we propose the use control techniques based on invariants to design a CZ gate with tunable-coupling transmons where $J(t)$ acts as the control. It is interesting to point out that the driving of the $\ket{11}$ state is achieved by the interaction with states out of the computational basis ($\ket{02}$ and $\ket{20}$, see Eq.\ \eqref{Heff}). Consequently, the designed protocols have to be particularly robust to avoid leakage.

The operation will be designed in three steps; {\itshape(i)} the coupling is turned on and engineered
from the initial value $J(0)=0$ to $J(T)=J_M$, {\itshape(ii)} the control remains constant during a waiting time $t_w$, {\itshape(iii)} the coupling is symmetrically switched off.
As result, the gate is implemented in a time $T_g=2T+t_w$, see Fig.\ \ref{fig:control}. The shape of the switch-on and off ramps will be determined by the non-adiabatic method used to design the operation.

If the two transmons are identical (same frequency and anharmonicity), the non-computational states $\ket{02}$ and $\ket{20}$ are degenerate, showing an avoided crossing behaviour when the interaction is ``on". If instead, the frequency of the second qubit is set to $\omega_b=\omega_a+\alpha_a$, the $\ket{11}$ and $\ket{20}$ states are brought to degeneracy while the $\ket{02}$ state is separated by an energy gap $|\alpha_a+\alpha_b|$. As result of such interaction, the $\ket{11}$ computational state acquires an adjustable phase compatible with the  performance of a CZ gate~\cite{Barends2014, Dicarlo2009}.
Under said condition, the effective matrix Hamiltonian becomes
\beq
  \label{HeffCZ}
  \hat H(t)  =\left( \begin{matrix}
      0 & 0 & 0 & 0 & 0 & 0 \\
      0 & \omega_a+\alpha_a & \widetilde{J}_1(t) & 0 & 0 & 0 \\
      0 & \widetilde{J}_1(t) & \omega_a & 0 & 0 & 0 \\
      0 & 0 & 0 & \Omega_2 & \widetilde{J}_2(t)& 0 \\
      0 & 0 & 0 & \widetilde{J}_2(t) & \Omega_1 & \widetilde{J}_3(t) \\
      0 & 0 &0 & 0 & \widetilde{J}_3(t) & \Omega_1
    \end{matrix}\right),
\eeq
where $\Omega_1=2\omega_a + \alpha_a$ and $\Omega_2=2 \omega_a+2\alpha_a+\alpha_b$. 
%
%

The dynamics are solved independently in each subspace. The $S_3$ can be approximately reduced to an effective 2x2 system, since the $\ket{02}$ state decouples from the
dynamics due to the $\Delta E=|\alpha_a+\alpha_b|\gtrsim 20J_M$ energy mismatch. As result the system Hamiltonian becomes
\beq
  \label{HeffCZ-2}
  \hat H(t)  \simeq\left( \begin{matrix}
      0 & 0 & 0 & 0 \\
      0 & \hat H_2(t) & 0 & 0 \\
      0 & 0 & \Omega_2 & 0 \\
      0 & 0 & 0 & \hat H_3(t)
    \end{matrix}\right).
\eeq
This means that the dynamics of $\{\ket{20},\ket{11}\}$ are approximately described by 
\beq
\label{H3}
\hat H_3(t)=(2\omega_a+\alpha_a)\mathbb{I}+\widetilde{J}_3(t)\sigma_1,
\eeq
where $\sigma_1$ represents the first Pauli matrix. The unitary transformation produced by Hamiltonian\ \eqref{H3} according to the path depicted in Fig.\ \ref{fig:control} is simply
\beq
\label{U3CZ}
\hat U_3(T_{g})=e^{-i(2\omega_a+\alpha_a) T_{g}}\exp\left[-i\hat\sigma_1(2\widetilde{J}_{3,T}+\widetilde{J}_3 (T) t_{w})\right],
\eeq
where $\widetilde{J}_{3,T}=\int_{0}^Tdt\widetilde{J}_3(t)$. Independently of the shape of the control $J(t)$, in order to recover the desired final state (times a phase), i.e., $\ket{11}\rightarrow e^{-i(2\omega_a+\alpha_a) T_g}(-\ket{11})$, the waiting time has to be properly adjusted to
\beq
\label{tw}
t_w=(\pi-2\widetilde{J}_{3,T})/\widetilde{J}_3(T).
\eeq
On the other hand, the $S_2$ subspace is governed by 
\beq
\label{H2}
\hat H_2(t)=(\omega_a+\frac{\alpha_a}{2})\mathbb{I}+\frac{\alpha_a}{2}\hat\sigma_3+\widetilde{J}_1(t)\hat\sigma_1,
\eeq
where $\sigma_1$ and $\sigma_3$ represent the first and third Pauli matrices. As at initial $t=0$ and final $t=T_g$ times the coupling is switched-off ($J=0$), the computational 
$\ket{01}$ and $\ket{10}$ states correspond to eigenstates of $\hat H_2$, thus an adiabatic evolution would drive this state frictionlessly restoring the original configuration at final time, but at expenses of a very long gate-time to avoid unwanted transitions. This limitation can be overcome by applying quasi-adiabatic quantum control techniques, such as the fast quasiadiabatic dynamics (FAQUAD) method~\cite{Martinez-Garaot2015}, or by inverse-engineering the control using invariants of motion~\cite{Torrontegui2013, Guery-Odelin2019}. Finally, the $S_1$ subspace is driven trivially.

\subsection{FAQUAD passage}
The fast quasiadiabatic dynamics method is a commonly used technique that aims at speeding up a given process while still making it as adiabatic as possible at all times. It relies on making the standard adiabaticity parameter constant such that the transition probability is equally delocalized during the whole process,
\beq
\label{FAQUAD}
\mu (t)\equiv\bigg\lvert\frac{\bra{+(t)}\dot{\hat H}_2\ket{-(t)}}{\left[E_+(t)-E_-(t)\right]^2}\bigg\rvert=\mu,
\eeq
where the dot represents the time derivative, $\ket{+(t)}$ and $\ket{-(t)}$ are the two instantaneous eigenstates of $\hat H_2(t)$ and $E_+(t)$ and $E_-(t)$ their respective eigenenergies. The process is considered to be adiabatic if $\mu\ll 1$. From Eq.\ \eqref{FAQUAD} it follows (see Appendix\ \ref{APadi})
\beq
\label{FAQUAD2}
\mu = \frac{\dot{\widetilde{J}_1}(t)\alpha_a}{\left[4\widetilde{J}_1^2(t)+\alpha_a^2\right]^{3/2}}.
\eeq
Equation\ \eqref{FAQUAD2} can be integrated with the initial and final conditions $J(0)=0$ and $J(T)=J_M$, which set a value for the integration constant and the adiabaticity parameter. The control is thus
\beq
\label{FAQUADcontrol}
\widetilde{J}_1(t)=\frac{-\alpha_a\widetilde{J}_1(T)t}{T\sqrt{\alpha_a^2+4\widetilde{J}_1^2(T)\left[1-\left(\frac{t}{T}\right)^2\right]}},
\eeq
which has the same profile for every value of the ramp time $T$, since $\widetilde{J}_1(t)$, as given by Eq.\ \eqref{FAQUADcontrol}, is a function of $t/T$. However, it can be seen that the adiabaticity parameter obtained with this control is inversely proportional to $T$ (see Appendix\ \ref{APadi}) meaning that shorter times lead to ``less adiabatic" protocols and a corresponding loss of fidelity.

\subsection{Invariants passage}

We can address the limitation of shorter operation times by inverse-engineering the control $\widetilde{J}_1(t)$ using invariants of motion~\cite{Torrontegui2013, Guery-Odelin2019}. 
The $\hat H_2$ Hamiltonian has SU(2) structure, $\hat H_2(t)=\sum_j^3h_j(t)\hat T_j$ where $h_j(t)$ are general controls of the Hamiltonian and $\hat T_j=\hat\sigma_j/2$  
are the Lie algebra generators satisfying $[\hat T_j,\hat T_k]=i\epsilon_{jkl}\hat T_l$, with $\epsilon_{jkl}$ the Levi-Civita tensor. Standard formalisms~\cite{Chen2011} would lead to the requirement of a time dependent anharmonicity $\alpha_a(t)$. Instead, due to the constrains imposed on the
controls of $\hat H_2$, i.e., $h_2=0$ and $h_3=\alpha_a$ $\forall t$, we will make use of the dynamical algebra~\cite{Torrontegui2014} to engineer the invariant of motion and, consequently, $\widetilde{J}_1(t)$.

Associated with a Hamiltonian, there are infinitely many time-dependent Hermitian invariants of motion $\hat I(t)$ that satisfy~\cite{Lewis1969}
\beq
\label{inva}
\frac{d\hat I}{dt}\equiv\frac{\partial\hat I(t)}{\partial t}-\frac{1}{i\hbar}[\hat H_2(t),\hat I(t)]=0.
\eeq
A wave function $\ket{\Psi(t)}$ which evolves 
with $\hat H_2(t)$ 
can be expressed as a linear combination of invariant modes~\cite{Lewis1969}
\beq
\label{LR}
\ket{\Psi(t)}=\sum_{n}c_ne^{i\beta_n(t)}\ket{\chi_n(t)}
\eeq
where the $c_n$ are constant, and the real phases $\beta_n$ fulfil
\beq
\label{LR-phase}
\hbar\dot\beta_n(t)=\bra{\chi_n(t)}i\hbar\partial_t-\hat H_2(t)\ket{\chi_n(t)},
\eeq
and are explicitly computed in Appendix\ \ref{APpha}. The eigenvectors $\ket{\chi_n(t)}$ of the invariant are assumed to form a complete set and satisfy 
$\hat I(t)|\chi_n(t)\rangle=\lambda_n|\chi_n(t)\rangle,$ with $\lambda_n$ the corresponding constant eigenvalues. 
Assuming that the invariant is also a member of the dynamical algebra, it can be written as $\hat I(t)=\sum_j^3f_j(t)\hat T_j$ 
where $f_j(t)$ are real, time-dependent functions.  
Replacing  the closed forms of $\hat H_2$ and $\hat I$
into Eq.\ \eqref{inva}, the  functions $h_j$ and $f_j$ satisfy 
\beq
\label{engineer}
\dot f_j(t)-\frac{1}{\hbar}\sum_{k}^{3}\sum_{l}^{3}\epsilon_{jkl}h_{k}(t)f_l(t)=0, \quad j=1,2,3.
\eeq
Usually, these coupled equations are interpreted as a linear system of ordinary differential 
equations for $f_j(t)$ when the $h_j(t)$ components of the Hamiltonian are known~\cite{Kaushal1981, Kaushal1993, Monteoliva1994, Maamache1995, Kaushal1997}.
Instead, we use here an inverse perspective, and consider them as an algebraic system to be solved for 
the $h_j(t)$, when the $f_j(t)$ are given~\cite{Torrontegui2014}. When we inverse engineer the controls, the Hamiltonian is given at initial and final times.
According to Eq.\ \eqref{LR} the dynamical wave function is transitionless driven (remember that $\{c_n\}$ are constant) through the eigenstates of the invariant, independently 
of the duration of the process, $T$. Note that this is exact, in contrast with a quasiadiabatic driving, where transitions from the instantaneous eigenstates of the Hamiltonian appear as soon as the process is shorten. To this end,
the invariant $\hat I(t)$ (or, equivalently, $f_1(t)$, see Appendix\ \ref{APinv}) is designed to drive, through its eigenvectors, the initial states of the Hamiltonian $\hat H_2(0)$ to the corresponding states 
of $\hat H_2(T)$~\cite{Chen2010, Torrontegui2013, Guery-Odelin2019}. This is ensured by imposing the “frictionless conditions" at the boundary times~\cite{Chen2010}
\beq
\label{BC}
 [\hat H_2(t_b ), \hat I (t_b )] = 0, \quad t_b=0,\ T,
\eeq
so $\hat H_2$ and $\hat I$ share eigenstates at these boundary times. 
 
The system of equations \eqref{engineer} is compatible if
the condition $\sum_j^3f_j^2(t)=c^2$ is fulfilled, with $c^2$ an arbitrary constant. Here, we choose $c^2=\sum_j^3h_j^2(0)=\alpha_a^2$ to make $\hat I(0)=\hat H_2(0)$. 
Under this condition, the system is invertible and has infinite solutions reflecting the fact that, associated with a Hamiltonian, there are infinite invariants and
vice-versa~\cite{Ibanez2012}. For the structure of $\hat H_2$, the coupling $\widetilde{J}_1(t)$ takes the form (see Appendix\ \ref{APinv})
\beq
\label{gt}
\widetilde{J}_1(t)=\frac{1}{2\sqrt{c^2-f_1^2-\left(\frac{\dot f_1}{\alpha_a}\right)^2}}\bigg(\frac{\ddot f_1}{\alpha_a}+\alpha_a f_1\bigg),
\eeq
where, according to \eqref{BC} and \eqref{gt}, $f_1\equiv f_1(t)$ satisfies the boundary conditions
\beq
\label{BC_f1}
f_1(t_b)=-h_1(t_b)\sqrt{\frac{c^2}{h_1^2(t_b)+\alpha_a^2}}, \quad \dot f_1(t_b)=\ddot f_1(t_b)=0,
\eeq
to ensure a perfect driving connecting the eigenstates at $\hat H_2(0)$ with the corresponding states at $\hat H_2(T)$.
For the ramp-up of the control, $h_1(0)=0$ and $h_1(T)=2\widetilde{J}_1(T)$, whereas $h_1(T+t_w)=2\widetilde{J}_1(T)$ and $h_1(T_g)=0$ when switching
the coupling off. We can then interpolate $f_1(t)$ with a polynomial ansatz $f_1(t)=\sum_{m=0}^Ma_mt^m$ and determine the $a_m$ coefficients from Eq.\ \eqref{BC_f1}. In the simplest interpolation, $M=5$ to satisfy all 6 boundary conditions, although additional constrains can be applied~\cite{Levy2018}.

The resulting gate using either this method or the previous one is equivalent to a CZ gate up to local rotations in each individual qubit (see Appendix\ \ref{EffectiveGate}). However, as it will be shown in the following section, the control based on invariants of motions leads to a better driving of the states of the $S_2$ subspace, enabling higher fidelity gates.

\section{Applications}\label{test}

%
%
%
%
\begin{figure}[t]
    \centering
    \includegraphics[width=0.9\linewidth]{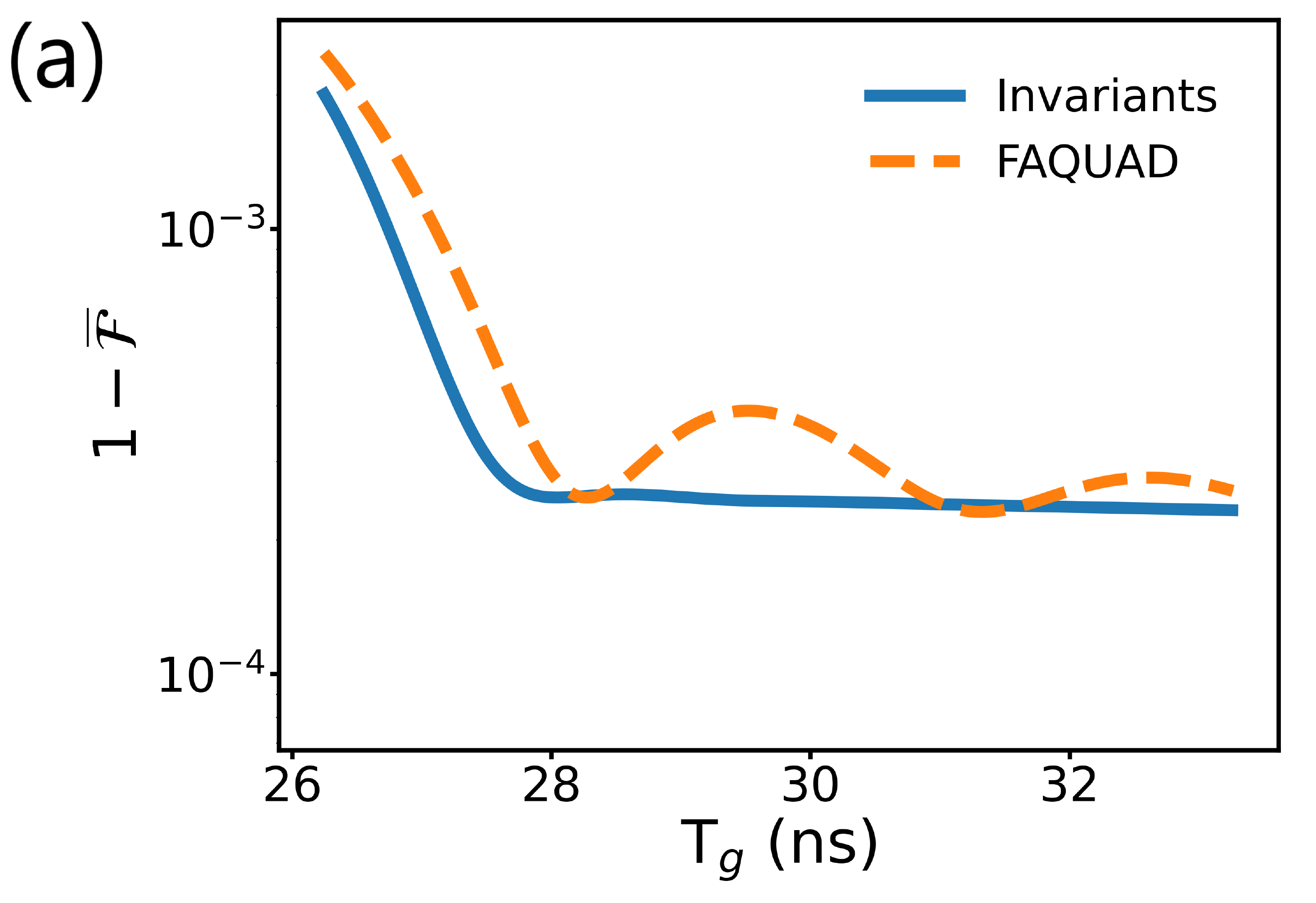}
    \includegraphics[width=0.9\linewidth]{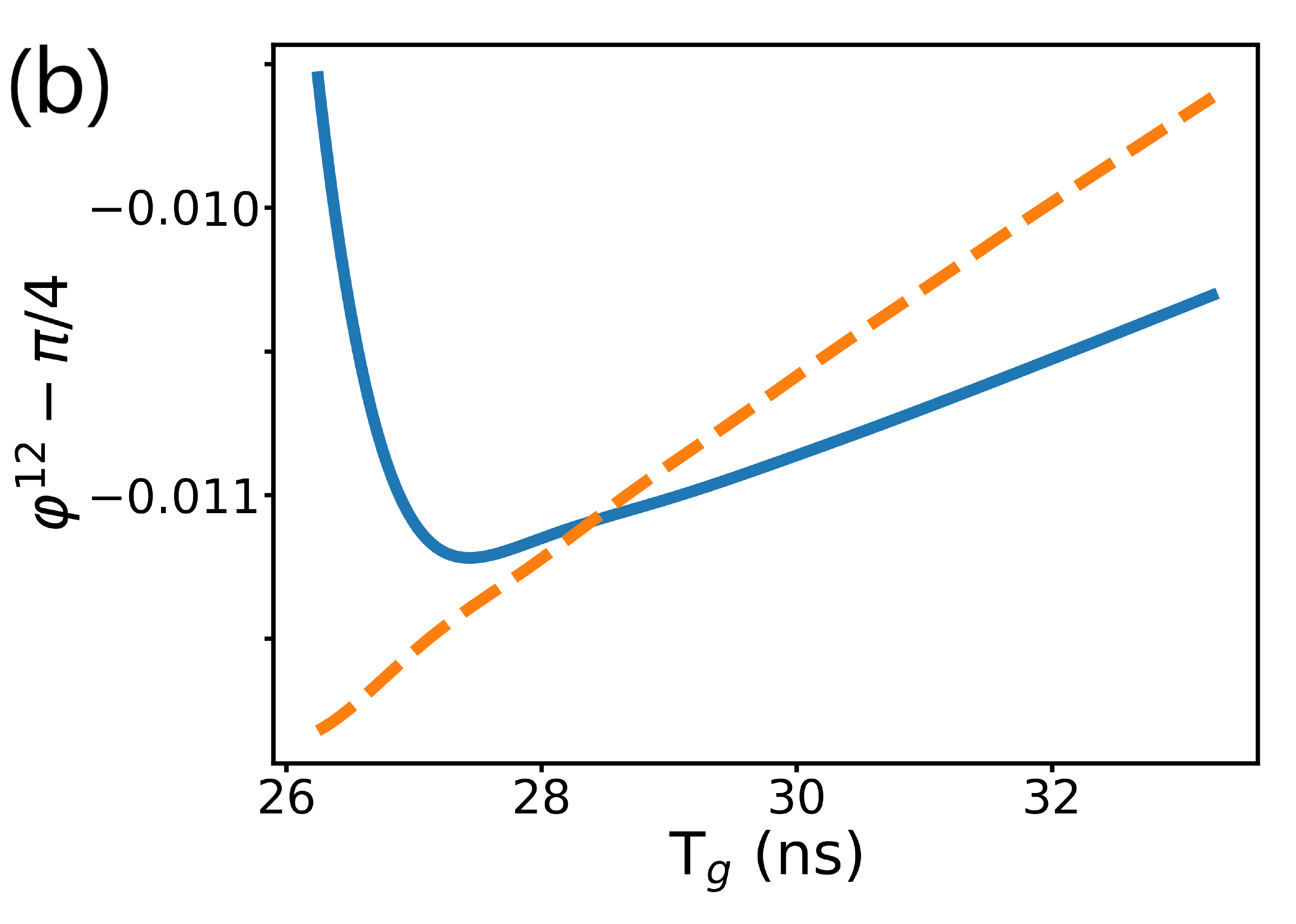}
    \includegraphics[width=0.9\linewidth]{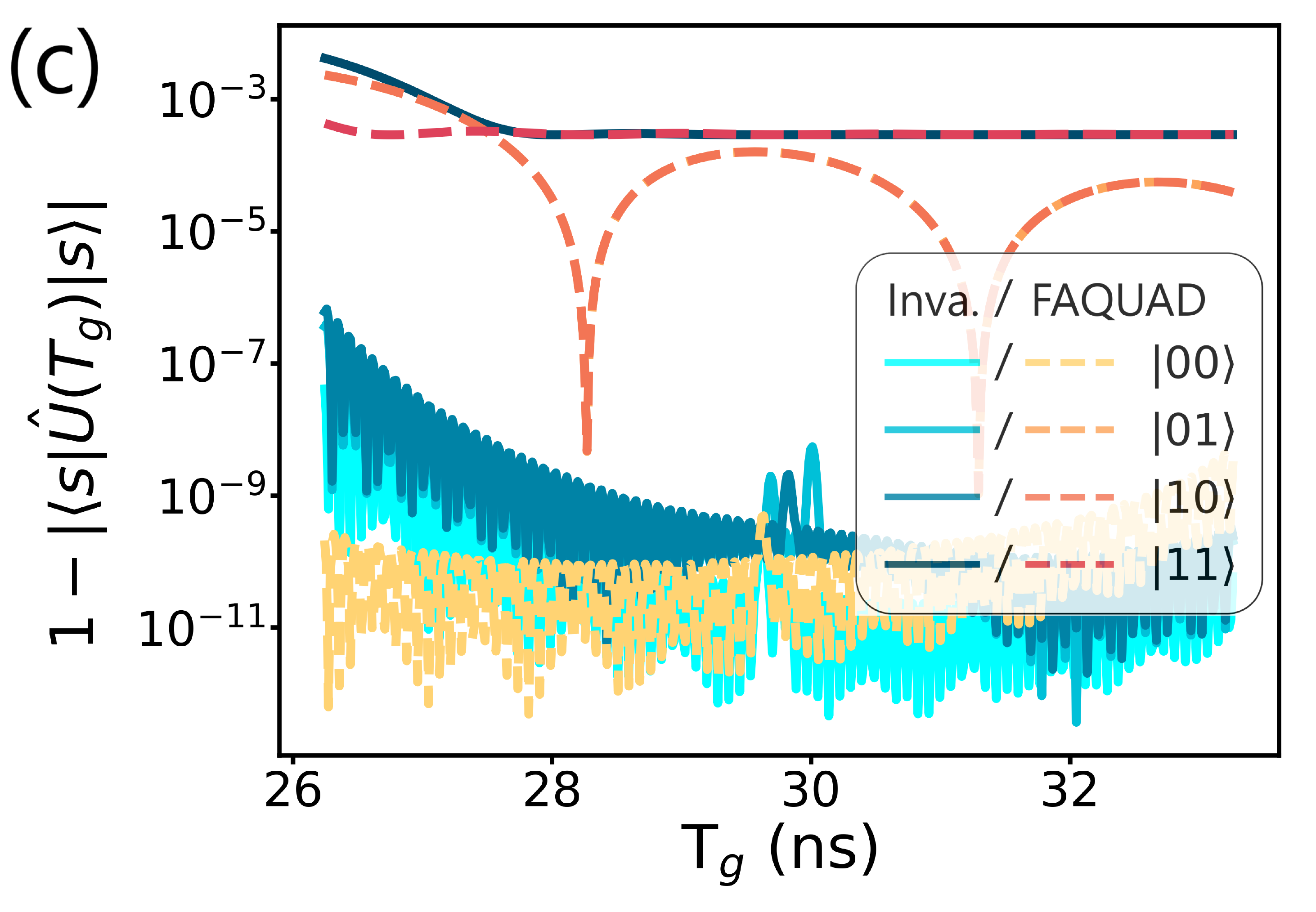}
  \caption{Performance of the invariants (solid blue line) and FAQUAD (orange dashed line) controls producing a CZ gate as a function of the total duration, with a ramp time ranging from 1 to 8 ns. (a) Average infidelity of the CZ gate. (b) Deviation of the entangling phase $\varphi^{12}$ from its expected value, $\pi/4$. (c) Average population loss of the computational basis states.}
  \label{fig:CZ_noD}
\end{figure}
%
%
%

In this section, we numerically simulate the system of two transmons using the full Hamiltonian given by \eqref{H0} and \eqref{H} in order to test the protocols designed for the reduced effective model. Due to the multiple approximations made through the derivations, such as imperfections in the reduced model and higher order energy corrections,
the obtained results are slightly different from the ideal CZ gate, even after rectifying local phases. The entangling phase, defined as %
\beq
  \label{phi12}
  \varphi^{12}=\frac{\phi_{00}-\phi_{01}-\phi_{10}+\phi_{11}}{4},
\eeq
where $\phi_{ij}$ is the phase accumulated by the $\ket{ij}$ state, might deviate from its expected value for a CZ gate, i.e., $\pi/4$, see Appendix\ \ref{EffectiveGate}. This difference, together with population losses in each state, implies a lower average gate fidelity, $\bar{\mathcal{F}}$, the figure of merit used in this paper to analyse the performance of the engineered controls. The average fidelity is calculated from the entanglement fidelity $\mathcal{F}_e$ as~\cite{Nielsen2002}
\beq
\label{fid}
\bar{\mathcal{F}}=\frac{N\mathcal{F}_e+1}{N+1},
\eeq
where N is the size of the Hilbert space ($N<\infty$ in a numerical simulation). The entanglement fidelity compares the final states resulting from the evolution $\hat U(T_g)$ of each computational state with the ideal operation $\hat U_{id}$ to be implemented. In our computations, we add an operation $\hat U^\dagger_{loc}$ that eliminates the effect of locally correctable phases, in a way that only the entangling phase is taken into account. Thus, the entanglement fidelity is defined as 
\beq
\mathcal{F}_e[\hat U_{id},\hat U(T_g)]=\Bigg\lvert\frac{1}{4}\sum_{s=1}^4\bra{s}\hat U^\dagger_{id}\hat U^\dagger_{loc}\hat U (T_g)\ket{s}\Bigg\rvert^2,
\eeq
with $\ket{s=1,2,3,4}=\{\ket{00},\ket{01},\ket{10},\ket{11}\}$.
%
%
%
%
%
%


\subsection{Uncorrected gates}
We now study the possibility of implementing the CZ gate and compare the performance of the gates obtained using the FAQUAD and invariants approaches. In Fig.\ \ref{fig:CZ_noD}(a) we show the average gate fidelity (or rather, the infidelity, defined as $1-\bar{\mathcal{F}}$) using both designs as a function of the total time taken by the operation. These curves are obtained by varying the ramp time $T$ from 1 to 8 nanoseconds and calculating the corresponding waiting time $t_w$ using Eq.\ \eqref{tw}. Both the FAQUAD and the invariant passages achieve fidelities over 99.9\%, with the invariant protocol giving slightly better results in general. The infidelity saturation at the same value, regardless of the applied control, is due to a common issue of both protocols: the driving of the $\ket{11}$ state is based on a coarse approximation, and corrections coming from a weak interaction with the $\ket{02}$ state lead to a loss of fidelity. 

The infidelity is the result of {\itshape(i)} the phase $\varphi^{12}$ deviating from its expected value, $\pi/4$, and {\itshape(ii)} the populations of states from the computational basis suffering from losses, such as undesired transfer to other states from the basis or even leakage outside of it, such that $\left|\bra{s}\hat U(T_g)\ket{s}\right|<1$. The phase deviation using the two kinds of control is shown in Fig.\ \ref{fig:CZ_noD}(b), while the population loss of each state is shown in Fig.\ \ref{fig:CZ_noD}(c). A common characteristic of controls based on shortcuts to adiabaticity, such as driving the states using dynamical invariants, is that they are minimally affected by leakage. This characteristic can be observed in the minimal population loss suffered by the $\ket{01}$ and $\ket{10}$ states when using the invariants passage. By contrast, the population loss of these states when using the FAQUAD passage represents a limiting factor in the achieved fidelity. The results shown in these figures suggest that both the phase deviation from $\pi/4$ and the population losses are substantial sources of error that shall be mitigated. In particular, the population of the $\ket{11}$ state is saturating the infidelity curves. Since the driving of the $\ket{11}$ state does not depend on the shape of $J(t)$, only on its integral, the saturation value is the same regardless of the protocol. As we will see, this saturation of the infidelity does not correspond to a fundamental limitation and can be attributed to approximations in the reduced model and a Stark-shift produced in the $S_3$ subspace by the interaction between $\ket{11}$ and $\ket{20}$. We shall introduce higher energy corrections and adjust our protocols to increase the average fidelity.

\subsection{Stark-shift corrected gates}
With the argument $|\alpha_a+\alpha_b|\gtrsim 20J_M$ to decouple the $\ket{02}$ state from the dynamics on the $S_3$ subspace, we found an analytical expression for the waiting time $t_w$, Eq.\ \eqref{tw}. However, this approximation needs to be examined in greater depth in order to improve the performance of the designed protocols. From \eqref{HeffCZ}, the $S_3$ subspace Hamiltonian after shifting the zero of energy becomes
\beq
  \hat H_3(t)  =\left( \begin{matrix}
      \alpha_a+\alpha_b & \widetilde{J}_2(t) & 0\\
      \widetilde{J}_2(t) & 0 & \widetilde{J}_3(t) \\
      0 & \widetilde{J}_3(t) & 0 
    \end{matrix}\right).
\eeq
%
%
%
%
\begin{figure}[t!]
    \centering
    \includegraphics[width=0.9\linewidth]{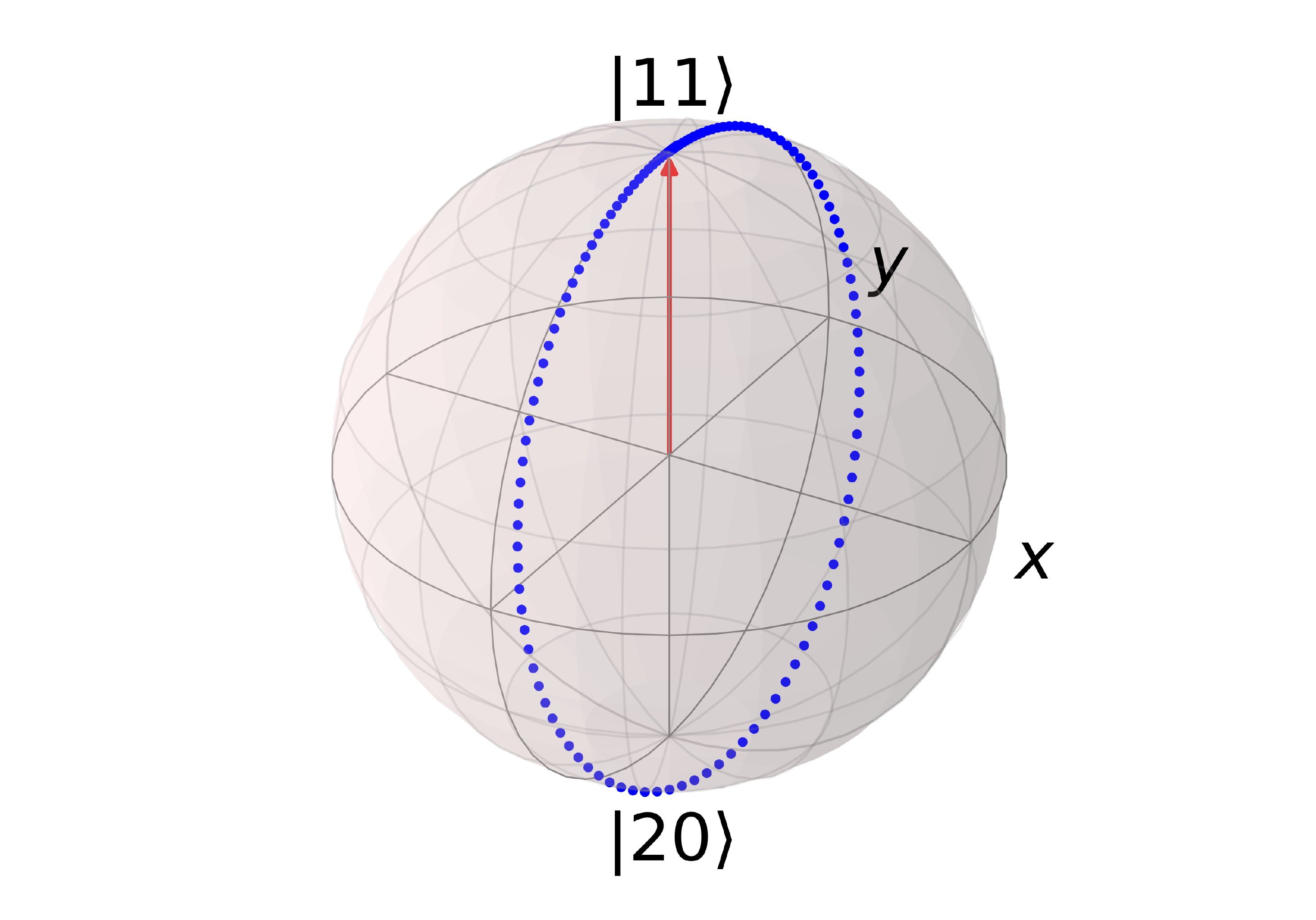}
  \caption{Visual representation of the evolution of the $\ket{11}$ state on the Bloch sphere under a Hamiltonian with the form of Eq.\ \eqref{H3eff}, with a magnified $\delta\Omega$ to perceive its effect. The $\ket{20}$ state is not attained, but due to the symmetry of the control protocols, the state can be brought back to $\ket{11}$ using an appropriate waiting time.}
  \label{fig:Bloch}
\end{figure}
%
%
%
In order to find the low-energy effective Hamiltonian for the $\ket{11}$ and $\ket{20}$ states, we apply the Schrieffer-Wolff transformation~\cite{Schrieffer1966} (see Appendix\ \ref{APswt} for the full derivation), leading to
\beq\label{H3eff}
\hat H_3^{ef}=\left( \begin{matrix}
	 \delta \Omega & \widetilde{J}_3+\delta \widetilde{J}_3 \\[6pt]
    \widetilde{J}_3+\delta \widetilde{J}_3 & 0 
    \end{matrix}\right),
\eeq
where the time dependence has been dropped for simplicity, and
\beq
\delta\Omega = \dfrac{\widetilde{J}_2^2(\alpha_a+\alpha_b)}{\widetilde{J}_3^2-(\alpha_a+\alpha_b)^2},
\eeq
\beq
\delta\widetilde{J}_3 =-\frac{1}{2} \dfrac{\widetilde{J}_2^2\widetilde{J}_3}{\widetilde{J}_3^2-(\alpha_a+\alpha_b)^2}.
\eeq
%
%
%
\begin{figure}[t!]
    \centering
    \includegraphics[width=0.9\linewidth]{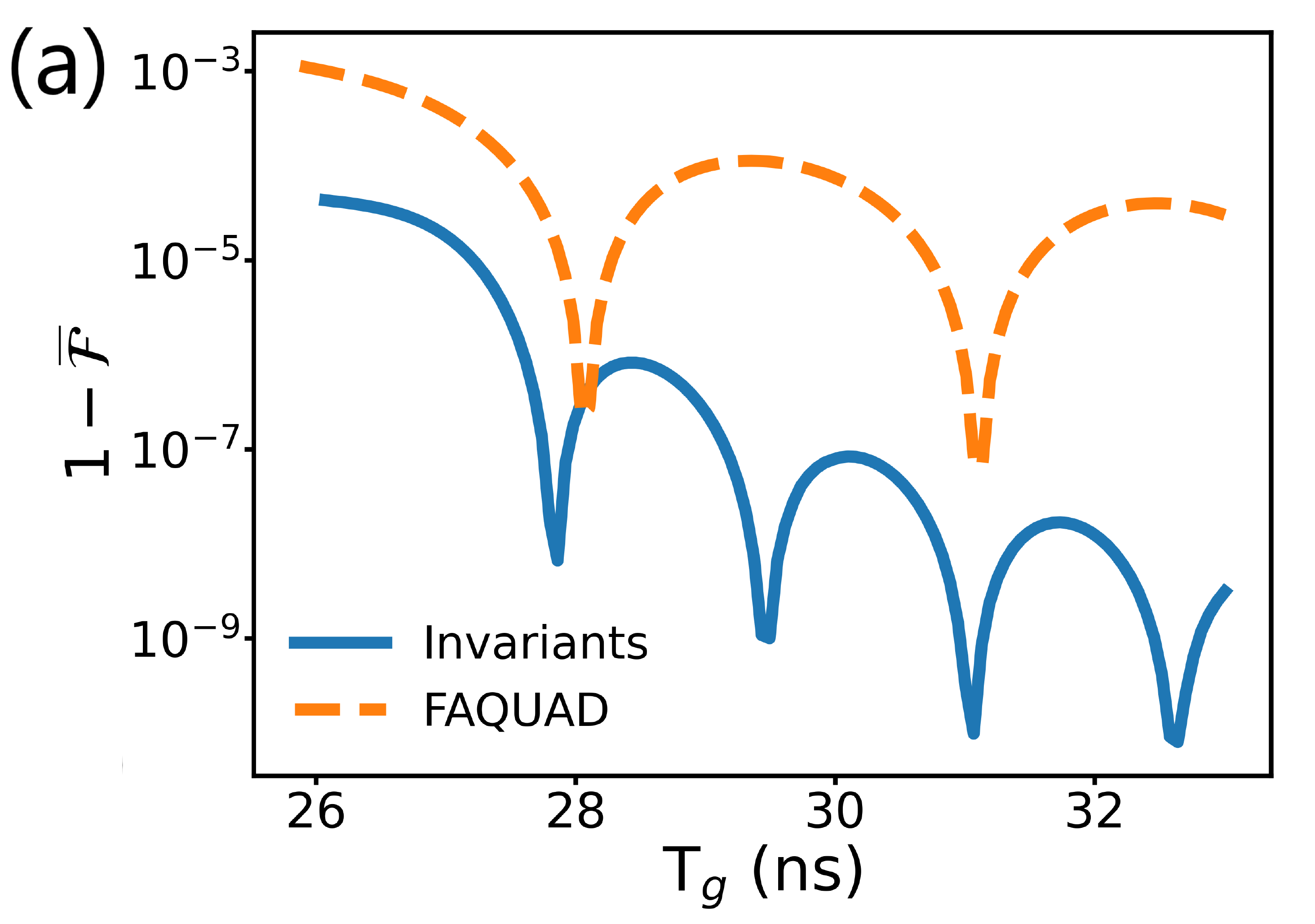}
    \includegraphics[width=0.9\linewidth]{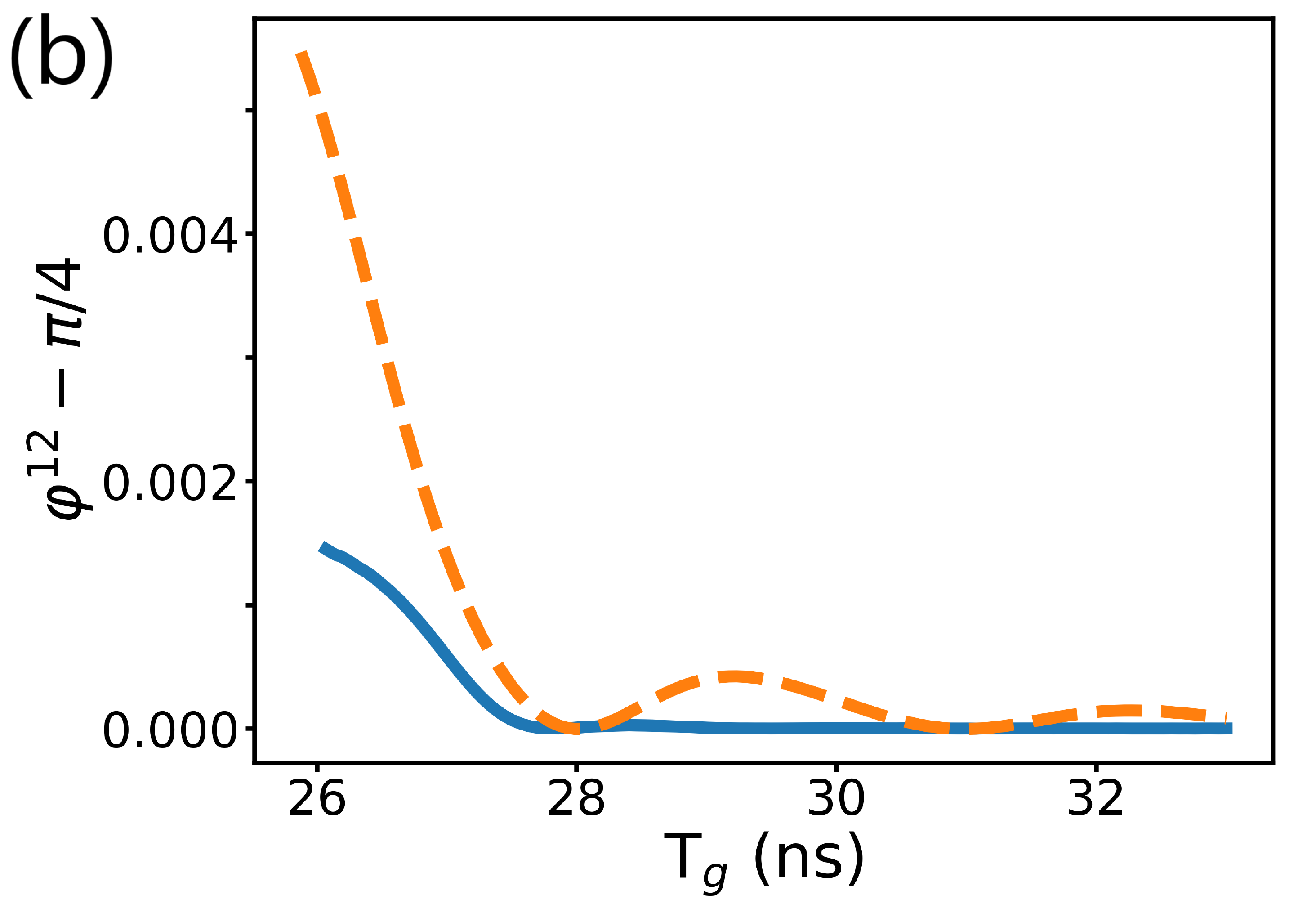}
    \includegraphics[width=0.9\linewidth]{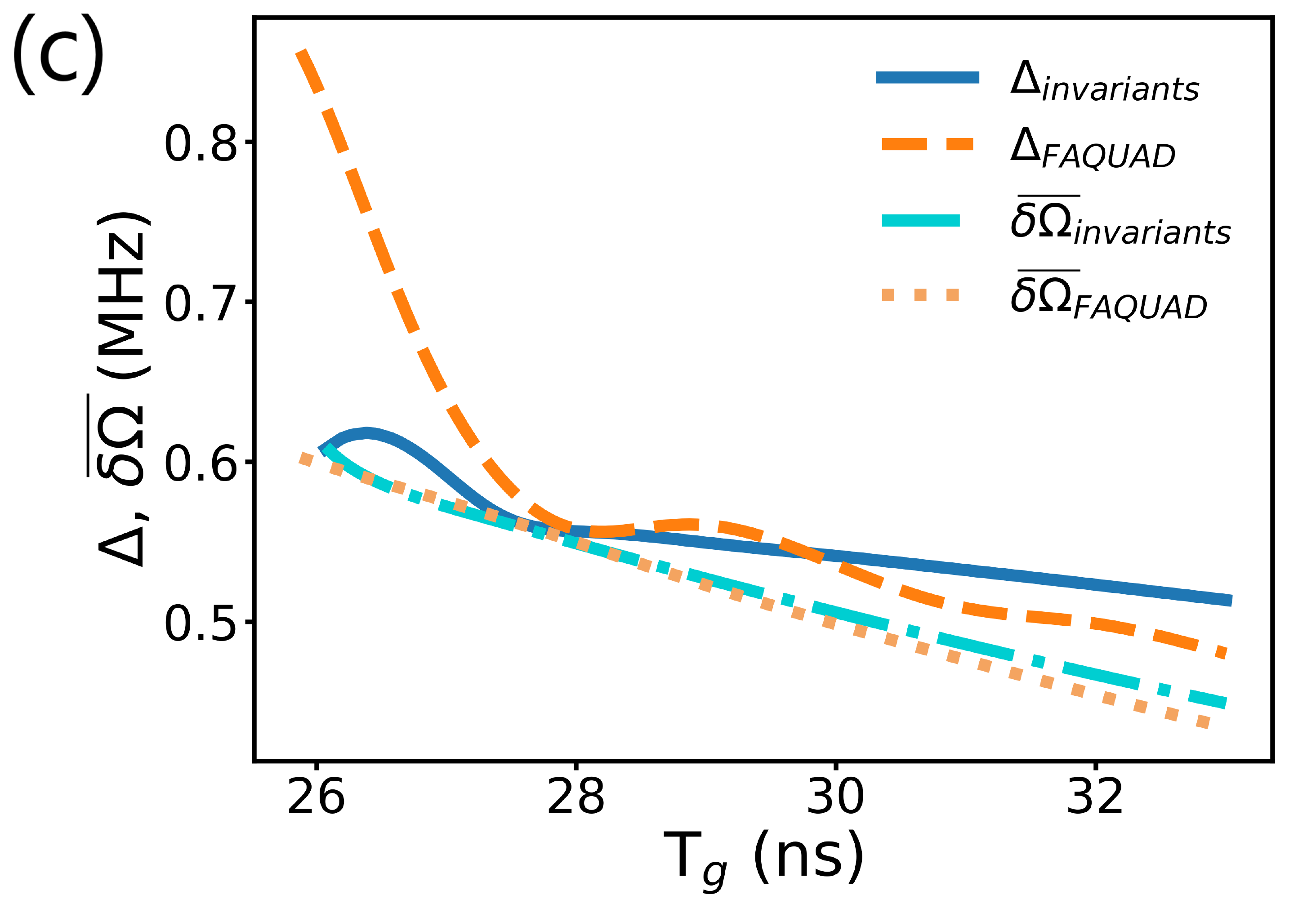}
  \caption{Optimized CZ gate with a detuning $\Delta$ in one of the transmon frequencies and an optimal waiting time. (a) Average infidelity of the gate as a function of the total duration, with a ramp time ranging from 1 to 8 ns. (b) Deviation of $\varphi^{12}$ from $\pi/4$ as a function of the gate duration. (c) Introduced detuning $\Delta$ and mean value of the Stark shift $\overline{\delta\Omega}$ as a function of the gate duration.}
  \label{fig:CZ_D}
\end{figure}
%
%
%
It is then clear that the interaction between the $\ket{11}$ and $\ket{02}$ states gives rise to a Stark shift $\delta \Omega$ in the energy of $\ket{11}$. The effect of $\hat H^{ef}_3$ acting on the $\ket{11}$ state can be visualized using the Bloch sphere with antipodal points corresponding to the pseudospin states $\ket{11}$ and $\ket{20}$. As it can be seen in this representation, the presence of $\delta \Omega$ (magnified in the Bloch sphere representation of Fig.\ \ref{fig:Bloch} for a clearer visualization) misaligns the rotation axis preventing a perfect $\ket{11}\leftrightarrow\ket{20}$ swap. As result, the Rabi oscillation amplitude attenuates and its period is shorten, since the rotational speed is increased by the $\hat \sigma_3$ term. This, added to the approximations of the reduced Hamiltonian model, produces a slight difference in the waiting time compared to Eq.\ \eqref{tw}, which explains the relatively high leakage of the $\ket{11}$ state in Fig.\ \ref{fig:CZ_noD}(c), independently of the followed protocol. Besides, the accumulated phase of the $\ket{11}$ state is also affected by the Stark shift, resulting in an entangling phase $\varphi^{12}\neq\pi/4$. As inferred from Fig.\ \ref{fig:CZ_noD}(b), corrections of this entangling phase by just mean of the waiting time would lead to very slow CZ gate implementations due to the $\sim\mu$s phase rate change. 

However, both the population loss of the state $\ket{11}$ and $\varphi^{12}$ can be simultaneously rectified by leaving the second qubit slightly detuned with respect to the resonance, i.e., $\omega_b = \omega_a+\alpha_a+\Delta$, where $|\Delta| \ll \widetilde{J}_2(T),\widetilde{J}_3(T)$. On the one hand, this constant detuning $\Delta$ sets the $\ket{11}$ state to rotate along the time dependent axis $(\widetilde{J}_3+\delta\widetilde{J}_3)\vec{u}_x+(\Delta+\delta\Omega)\vec{u}_z/2$, see Appendix\ \ref{APswt}. Nevertheless, the symmetry of the designed protocols (see Fig.\ \ref{fig:control}) allows us to always perfectly recover the $\ket{11}$ state with an appropriate waiting time, independently of the $\Delta$ value. Furthermore, a numerical adjustment of the waiting times also rectifies the population loss produced by the energy deviations of the effective model with respect to the complete Hamiltonian. On the other hand, a fine tune of $\Delta$ allows the rectification of $\varphi^{12}$, even if $\delta \Omega$ is in fact time-dependent. Notice that this detuning also affects the $S_2$ subspace, and therefore the designed protocols must be corrected by substituting $\alpha_a\rightarrow \alpha_a+\Delta$ in Eqs.\ \eqref{FAQUADcontrol} and \eqref{gt}. As a matter of fact, if $|\Delta|\lll|\alpha_a|$, the controls barely change. 

In Fig.\ \ref{fig:CZ_D} we show the results of using corrected protocols in which a small detuning $\Delta$ is introduced in the frequency of one of the transmons ($\omega_b = \omega_a+\alpha_a+\Delta$), and both the waiting time and this detuning are found numerically to minimize the gate infidelity -- computed from Eq.\ \eqref{fid} -- for a range of ramp times $T$ from 1 to 8 ns. Figure \ref{fig:CZ_D}(a) shows a drastic improvement of the gate fidelity, especially in the protocol designed using invariants of motion, where the fidelity is not limited by the population loss of states $\ket{01}$ and $\ket{10}$. This figure shows the great potential of nonadiabatic protocols over adiabatic processes, in which $T\rightarrow \infty$ is required for a truly faultless transition between the eigenstates of the instantaneous Hamiltonian. In Fig.\ \ref{fig:CZ_D}(b) we show how the entangling phase deviation is also attenuated by optimally detuning the two transmons, which leaves the population losses as the main source of errors. The optimal detuning is shown in Fig.\ \ref{fig:CZ_D}(c) as a function of the gate time. Notice that, even though $\delta \Omega$ is positive for the chosen range of parameters, the detuning is also positive. This may seem counter-intuitive at first from Eq.\ \eqref{H3eff}, but it is actually compatible with all our derivations. The detuning not only affects the phase of the $\ket{11}$ state, $\phi_{11}$, but also $\phi_{01}$ and $\phi_{10}$ by approximately the same amount. Thus, a correction in the $\phi_{11}$ phase leads to approximately the same correction in $\varphi^{12}$ with opposite sign, see Eq.\ \eqref{phi12}. The detuning is about the same order and sign as the mean value of $\overline{\delta \Omega}$ over the whole process, which confirms our assumption $|\Delta|\lll|\alpha_a|$.

\section{Conclusions}\label{conclusions}
To sum up, we demonstrate the possibility of implementing a CZ gate using a control protocol that exploits dynamical invariants of motion to drive the states frictionlessly in a system of two transmons and a tunable coupler. The designed control takes advantage of having the coupling strength at its maximum value during most of the gate time, accelerating the processes close to their speed limits. This method has been compared with the adiabatic process, showing that the method based on invariants leads to an overall better performance for any gate time. The protocol was found analytically using a simplified effective Hamiltonian that describes the lowest six energy levels of the two-transmon system. We then numerically adjusted the waiting time and the destination frequency of the qubits -- slightly out of the avoided crossing between $\ket{11}$ and $\ket{02}$-- to correct higher order errors coming from the Stark shift produced by the $\ket{20}$ state. With these corrections, the invariants method achieves infidelities 2 to 3 orders of magnitude lower than the adiabatic protocol.

The demonstration of such a high fidelity CZ gate shows the viability of the theory of dynamical invariants to construct fast diabatic gates in a tunable-coupling qubit architecture. Short gate times, low losses, and reported greater robustness to decoherence make control protocols designed using the theory of dynamical invariants superior candidates for implementing fast and high-fidelity quantum gates in existing setups.
\acknowledgments
We acknowledge financial support from the Spanish Government through PGC2018-094792-B-I00 (MCIU/AEI/FEDER,UE), CSIC Research Platform PTI-001, and by Comunidad de Madrid-EPUC3M14 and CAM/FEDER Project No. S2018/TCS-4342 (QUITEMAD-CM). H.E. acknowledges the Spanish Ministry of Science,
Innovation and Universities for funding through the FPU
program (FPU20/03409). E.T. acknowledges the Ramón y Cajal program (RYC2020-030060-I).
\appendix
\section{Design of FAQUAD passage}\label{APadi}
After shifting the zero of energy, Hamiltonian $\hat H_2(t)$ from Eq.\ \eqref{H2} can be written in matrix form as
\beq
  \hat H_2(t)  =\left( \begin{matrix}
      \dfrac{\alpha_a}{2} & \widetilde{J}_1(t)  \\[6pt]
      \widetilde{J}_1(t) & -\dfrac{\alpha_a}{2}
    \end{matrix}\right).
\eeq
The instantaneous eigenstates of this Hamiltonian are
\beq
\ket{+(t)}=
\left(\begin{array}{c} 
\cos\frac{\theta(t)}{2} \\[6pt]
-\sin\frac{\theta(t)}{2}  
\end{array}\right),
\hspace{0.3cm}
\ket{-(t)}=
\left(\begin{array}{c} 
 \sin\frac{\theta(t)}{2} \\[6pt]
 \cos\frac{\theta(t)}{2}
\end{array}\right),
\eeq
where $\theta (t)=\arctan\frac{\alpha_a}{2\widetilde{J}_1(t)} $. Thus, we find
\beq
\bra{+(t)}\dot{\hat H}_2\ket{-(t)}=\dot{\widetilde{J}_1}(t)\cos\theta= \frac{\dot{\widetilde{J}_1}(t)\alpha_a}{2\sqrt{\left(\frac{\alpha_a}{2}^2\right)+\widetilde{J}^2_1(t)}}.
\eeq
The energy associated with each eigenstate is
\beq
E_\pm(t)=\pm\sqrt{\left(\frac{\alpha_a}{2}^2\right)+\widetilde{J}^2_1(t)},
\eeq
and using Eq.\ \eqref{FAQUAD}, we get
\beq
\mu = \frac{\dot{\widetilde{J}_1}(t)\alpha_a}{\left[4\widetilde{J}_1^2(t)+\alpha_a^2\right]^{3/2}}.
\eeq
Integrating both sides of the equation, we find
\beq
\mu t + c_0= \frac{-\alpha_a}{4\left[4\widetilde{J}_1^2(t)+\alpha_a^2\right]^{1/2}},
\eeq
where $c_0$ is an integration constant. The previous equation can be inverted into
\beq
\widetilde{J}_1(t)= \frac{-\alpha_a}{2}\sqrt{\frac{1}{16\left(\mu t+c_0\right)^2}-1}.
\eeq
Finally, using the boundary conditions on $\widetilde{J}_1(t)$ at $t=0$ and $t=T$, we find the value of $c_0$ and $\mu$
\beq
c_0 = \frac{\alpha_a}{4},\hspace{0.5cm}\mu = -\frac{2 \widetilde{J}_1(T)}{T\alpha_a}\frac{1}{\sqrt{16 \widetilde{J}^2_1(T)+ \alpha_a^2}}.
\eeq
Putting it altogether leads to
\beq
\widetilde{J}_1(t)=\frac{-\alpha_a\widetilde{J}_1(T)t}{T\sqrt{\alpha_a^2+4\widetilde{J}_1^2(T)\left[1-\left(\frac{t}{T}\right)^2\right]}}.
\eeq

\section{Inverse engineering of the controls}\label{APinv}

The set of equations \eqref{engineer} can be represented matricially
\beq
\label{sisSU2}
\left(\begin{array}{c} 
\dot f_1 \\
\dot f_2   \\
\dot f_3 
\end{array} \right)=
\underbrace{\frac{1}{\hbar}\left(\begin{array}{ccc} 
0&   f_3 & -f_2  \\
-f_3 &   0& f_1 \\
f_2      &  -f_1 & 0 
\end{array} \right)}_{=\mathcal{A}}
\left(\begin{array}{c} 
h_1  \\
h_2\\
h_3
\end{array} \right).
\eeq
As $\mathcal{A}=-\mathcal{A}^{\dagger}$ is a real antisymmetric matrix and with  odd dimensionality, the
eigenvalues are conjugate pure imaginary pairs, and zero, $a^{(0)}=0$, $a^{(1)}=-i\sqrt{\gamma}/\hbar$, and $a^{(2)}=i\sqrt{\gamma}/\hbar$ 
with $\gamma=f_1^2+f_2^2+f_3^2$. As $a^{(0)}=0$ then $\det(\mathcal{A})=0$, there is no inverse matrix $\mathcal{A}^{-1}$. However, we can still invert
the system, for example using Gauss elimination, to reduce the system to an equivalent one with the same solutions applying elementary operations. 
These are the multiplication of a row by a non-zero scalar,
the interchange of columns or rows, and the addition to a row of the multiple of a different one.
The augmented matrix of \eqref{sisSU2} is
\beq
{\left(\begin{array}{cccc} 
-f_3 &   0 & f_1  &\hbar\dot f_2\\
0 &   f_3&- f_2 &\hbar\dot f_1\\
f_2      & -f_1 & 0 &\hbar\dot f_3
\end{array} \right)}.
\eeq
After some algebra the system can be written as a lower triangular matrix,
\beq
{\left(\begin{array}{cccc} 
-f_3 &   0 & f_1  &\hbar\dot f_2\\
0 &   f_3 & -f_2  &\hbar\dot f_1\\
0     &  0 & 0 &\frac{\hbar(f_1\dot f_1+f_2\dot f_2+f_3\dot f_3)}{f_3}
\end{array} \right)}.
\eeq
This system is compatible and has infinite solutions if and only if $f_1\dot f_1+f_2\dot f_2+f_3\dot f_3=0$ 
or, equivalently,  $f_1^2+f_2^2+f_3^2=c^2$. The solutions satisfy
\beqa
\hbar\dot f_1&=&f_3h_2-f_2h_3,
\nonumber\\
\hbar\dot f_2&=&-f_3h_1+f_1h_3,
\eeqa
or in a compact form
\beq
\label{hs}
h_i=-\hbar{\cal{E}}_{ijk}\frac{\dot f_j}{f_k}+\frac{f_i}{f_k}h_k,
\eeq
with all indices $i,\, j,\, k$ different. $h_k(t)$ is considered a free function chosen for convenience,
for example, making it zero if we want to cancel the $h_k$ control of the Hamiltonian.

Our aim is to find $\hat H_2(t)=\frac{\alpha_a}{2}\hat\sigma_3+\widetilde{J}_1(t)\hat\sigma_1$ so that the ground and excited
states of $\hat H_2(0)$ become the ground and excited states of $\hat H_2(T)$
in an arbitrary time $T$, up to phase factors, in such a way
that $h_2(t) = 0, h_3(t)=\alpha_a$ $\forall t$. Choosing $(i,j,k)=(3,1,2)$ in Eq.\ \eqref{hs} and using $\gamma=c^2$ we can express $f_3$ and $f_2$ in terms of $f_1$,
\beqa
\label{eqF}
f_2&=&\frac{\dot f_1}{\alpha_a},
\nonumber\\
f_3&=&\sqrt{c^2-f_1^2-\frac{\dot f_1^2}{\alpha_a^2}}.  
\eeqa
Substituting this in the other equation of Eq.\ \eqref{hs},  with $(i,j,k)=(1,3,2)$, 
\beq
\label{control}
h_1(t)=\frac{1}{\sqrt{c^2-f_1^2-\frac{\dot f_1^2}{\alpha_a^2}}}\bigg(\frac{\ddot f_1}{\alpha_a}+\alpha_a f_1\bigg),
\eeq
leading to $\widetilde{J}_1(t)=h_1(t)/2$.

The ``frictionless conditions'' $[\hat H(t_b), \hat I(t_b)]=0$ for a closed Lie algebra of $\hat H$ and $\hat I$ can be reformulated as
\beq
\sum_{j,k,l}^{N}\epsilon_{jkl}h_k(t_b)f_l(t_b)\hat T_j=0. 
\eeq
Since the $\hat T_j$ generators are independent, the coefficients must satisfy
\beq
\sum_{k,l}^{3}\epsilon_{jkl}h_k(t_b)f_l(t_b)=0, \quad j=1,\dots,3,\quad t_b=0,T.
\eeq
For $\hat H_2(t)$, having $h_2(t)=0$ and $h_3(t)=\alpha_a$, these conditions read $f_2(t_b)=0$ and $\alpha_a/h_1(t_b)=f_3(t_b)/f_1(t_b)$ or, equivalently with the help of \eqref{eqF},
\beq
\label{co1}
f_1(t_b)=-h_1(t_b)\sqrt{\frac{c^2}{h_1^2(t_b)+\alpha_a^2}},\quad \dot f_1(t_b)=0.
\eeq
In addition, from Eq.\ \eqref{control} at the boundary times $t_b$,
\beq
\label{co2}
\ddot f_1(t_b)=0.
\eeq
We then interpolate $f_1(t)$ with a simple polynomial $f_1(t)=\sum_{j=0}^5a_jt^j$ where the $a_j$ coefficients are determined from Eqs.\ \eqref{co1} and \eqref{co2},
or following some more sophisticated approach, e. g. to optimize some additional constraint,  
and construct the control $\widetilde{J}_1(t)$ using Eq.\ \eqref{control}.

\section{The Lewis-Riesenfeld phase}\label{APpha}
During the ramp-up and down the Lewis-Riesenfeld phase $\beta_n$ accumulated by the states, see Eq.\ \eqref{LR-phase} is fully determined by the ramp time $T$ and the particular
election of the invariant auxiliary function $f_1(t)$.
Since the ramp-up and down processes are symmetric, meaning that the control at $t_w+T+t$ is identical to itself at $T-t$, the term coming from the time derivative in the Lewis-Riesenfeld phase has the opposite sign at each ramp, meaning that this term gets cancelled when the whole operation is performed. Therefore,
\beq 
  \label{LR-totphase}
  \beta^{tot}_n\equiv\beta^{up}_n+\beta^{down}_n=-2\int_0^T dt \bra{\chi_n(t)}\hat H(t)\ket{\chi_n(t)}.
\eeq
The eigenvectors and eigenvalues of the invariant $\hat I(t)=\sum_{j}^3f_j(t)\frac{\hat\sigma_j}{2}$ with $\sum_{j}^3f_j^2(t)=c^2$ acting on the $S_2={\ket{01},\ket{10}}$ subspace are
\begin{align}
\label{eigenCZ}
\ket{\chi_{\pm}}&=\frac{1}{\sqrt{2c(c\pm f_3)}}\left[\left(f_3\pm c\right)\ket{01}+\left(f_1+if_2\right)\ket{10}\right],\nonumber\\
\lambda_{\pm}&=\pm\frac{c}{2},
\end{align}
with $\ket{\chi_{+}}=\ket{\chi_{\ket{10}}}$ and $\ket{\chi_{-}}=\ket{\chi_{\ket{01}}}$. Replacing \eqref{eigenCZ} into Eq.\ \eqref{LR-totphase} with $h_1(t)=2\tilde J_1(t)$, $h_2(t)=0$ and $h_3(t)=\alpha_a$, we finally find 
\beq 
  \beta^{tot}_{\pm}=\mp\frac{2}{|\alpha_a|}\int_0^Tdt\left(2f_1\widetilde{J}_1+f_3\frac{\alpha_a}{2}\right).
\eeq
\section{Effective gate}\label{EffectiveGate}
In principle, due to different phases accumulated by each of the computational states, with the described procedure we do not get a CZ gate, but instead a transformation that, acting on the computational basis, is given by the unitary matrix
\beq
 \label{UCZ}
 \hat U = \left( \begin{matrix}
     e^{i\phi_{00}} & 0 & 0 & 0 \\
     0 & e^{i\phi_{01}} & 0 & 0  \\
     0 & 0 & e^{i\phi_{10}} & 0  \\
     0 & 0 & 0 & e^{i\phi_{11}} 
   \end{matrix}\right).
\eeq
In fact, this would only correspond to a CZ gate if $\phi_{00}=\phi_{01}=\phi_{10}=0$ and $\phi_{11}=\pi$. An alternative way of writing Eq.\ \eqref{UCZ} is
\beq
  \label{UCZ2}
  \hat U= \exp[i(\varphi^0+\varphi^1\hat\sigma_3\otimes\mathbb{I}+\varphi^2\mathbb{I}\otimes\hat\sigma_3+\varphi^{12}\hat\sigma_3\otimes\hat\sigma_3)].
\eeq
$\varphi^0$ represents a global phase, while $\varphi^1$ and $\varphi^2$ are local phases that can be corrected at each qubit using one-qubit gates. $\varphi^{12}$ is the entangling phase resulting from the interaction between the two qubits. Each of these phases can be written in terms of $\{\phi_{ij}\}$. In particular, the entangling phase,
\beq
  \varphi^{12}=\frac{\phi_{00}-\phi_{01}-\phi_{10}+\phi_{11}}{4}.
\eeq
Equations \eqref{UCZ2} and \eqref{phi12} reveal that the unitary transformation given by Eq.\ \eqref{UCZ} can be turned into a CZ gate by applying additional one qubit gates only if $\varphi^{12}=\pi/4$.

Let us explicitly calculate the entangling phase. According to the Hamiltonian \eqref{HeffCZ} the state $\ket{00}$ is left completely unchanged ($\phi_{00}=0$). Independently of the type of passage, FAQUAD or invariants, the states $\ket{01}$ and $\ket{10}$ acquire the following phases
\beq
\phi_{01}=-(\omega_a+\frac{\alpha_a}{2})T_g-\Omega't_w-2\int_0^T dt \langle\hat H_2(t)\rangle_{\ket{01}},
\eeq
\beq
\phi_{10}=-(\omega_a+\frac{\alpha_a}{2})T_g+\Omega't_w-2\int_0^T dt \langle\hat H_2(t)\rangle_{\ket{10}}, 
\eeq
where $\Omega'=\sqrt{(\alpha_a/2)^2+\widetilde{J}_1(T)^2}$, and $\langle\hat H_2(t)\rangle_{\ket{nm}}$ is the expectation value of the $S_2$ Hamiltonian at time $t$ when the state is initially $\ket{nm}$. This expectation value depends on the design of the control, since in the FAQUAD passage the states evolves approximately as the instantaneous eigenstates of $\hat H_2(t)$, while in the invariants passage, they coincide with the eigenstates of $\hat I(t)$. However, in both cases $\langle\hat H_2(t)\rangle_{\ket{01}}+\langle\hat H_2(t)\rangle_{\ket{10}}=0$, cancelling their contribution to the entangling phase. Finally, from Eq.\ \eqref{U3CZ}, the state $\ket{11}$ acquires a phase
\beq\label{phi11}
\phi_{11}=-(2\omega_a+\alpha_a)T_g+\pi,
\eeq
Thus, inserting these results into Eq.\ \eqref{phi12} we find that $\varphi^{12}$ is precisely equal to $\pi/4$. Consequently, the designed protocols produce a transformation which, after local rotations on each of the qubits, is equivalent to a CZ gate.
\section{Schrieffer-Wolff transformation}\label{APswt} 
The Schrieffer–Wolff transformation is a unitary transformation used to diagonalize a given Hamiltonian to first perturbative order in the interaction. It is often used to project out the high (low) energy excitations of a given quantum many-body Hamiltonian in order to obtain an effective low (high) energy model. The transformation is conventionally written as
\beq
\hat H'=e^{\hat S}\hat He^{-\hat S},
\eeq
where $\hat S$ is the generator of the transformation and $\hat H$ is a Hamiltonian that can be written as
\beq
\hat H=\hat H_0+\hat V,
\eeq
with $\hat H_0$ being block-diagonal and $\hat V$ purely off-diagonal in the eigenbasis of $\hat H_0$. In our particular case, the starting point is the $S_3$ subspace Hamiltonian that corresponds to the block of Eq.\ \eqref{HeffCZ}
\beq
  \hat H_3(t)  =\left( \begin{matrix}
      \alpha_a+\alpha_b+2\Delta & \widetilde{J}_2(t) & 0\\
      \widetilde{J}_2(t) & \Delta & \widetilde{J}_3(t) \\
      0 & \widetilde{J}_3(t) & 0 
    \end{matrix}\right),
\eeq
where we have included the second qubit detuning $\omega_b = \omega_a+\alpha_a+\Delta$ and the zero energy has been shifted. We chose
\beq\label{SWH0}
  \hat H_0  =\left( \begin{matrix}
      \alpha_a+\alpha_b+2\Delta & 0 & 0\\
      0 & \Delta & \widetilde{J}_3 \\
      0 & \widetilde{J}_3 & 0 
    \end{matrix}\right),
\eeq
    and
\beq\label{SWV}
    \hat V  =\left( \begin{matrix}
      0 & \widetilde{J}_2 & 0\\
      \widetilde{J}_2 & 0 & 0 \\
      0 & 0 & 0 
    \end{matrix}\right),
\eeq
with the time-dependence notation dropped for simplicity.
The transformation can be expanded in $\hat S$ using the Baker-Campbell-Haussdorf formula
\beq
\label{SWcondition}
\hat H'=\hat H_0+\hat V+[\hat S,\hat H_0]+[\hat S,\hat V]+\frac{1}{2}[\hat S,[\hat S,\hat H]]+\cdots.
\eeq
The Hamiltonian can then be made diagonal to first order in $\hat V$ by choosing the generator $\hat S$ such that
\beq
[\hat S,\hat H_0]=-\hat V,
\eeq
so that the off-diagonal terms are cancelled to the first order in the perturbation. The difficult step is the computation of the generator of the Schrieffer-Wolff transformation. The method presented in~\cite{Rukhsan2019} to calculate the generator starts by calculating the commutator $[\hat H_0,\hat V]$ and replace every non-zero matrix element of the commutator by undetermined coefficients. Then, those coefficients are 
computed using Eq.\ \eqref{SWcondition}. Using from the commutator of \eqref{SWH0} and \eqref{SWV}, the generator takes the form
\beq
\hat S =\left( \begin{matrix}
0 & a_1 & a_2\\
-a_1 & 0 & 0 \\
-a_2 &0 & 0
\end{matrix}\right).
\eeq
Using the condition \eqref{SWcondition}, the coefficients are found to be
\beq
a_1 = \dfrac{\widetilde{J}_2(\alpha_a+\alpha_b+2\Delta)}{-\widetilde{J}_3^2+(\alpha_a+\alpha_b+\Delta)(\alpha_a+\alpha_b+2\Delta)}, 
\eeq
\beq
a_2 = \dfrac{\widetilde{J}_2\widetilde{J}_3}{\widetilde{J}_3^2-(\alpha_a+\alpha_b+\Delta)(\alpha_a+\alpha_b+2\Delta)}.
\eeq
Finally, the effective Hamiltonian is calculated using the first order of the BCH formula from the exponential expansion,
\beq
\hat H' =\hat H_0+\frac{1}{2}\left[\hat S,\hat V\right],
\eeq
which gives
\beq
\hat H'=\left( \begin{matrix}
      \alpha_a+\alpha_b+2\Delta-\delta \Omega & 0 & 0\\[6pt]
      0 &\Delta + \delta \Omega & \widetilde{J}_3+\delta \widetilde{J}_3 \\[6pt]
      0 & \widetilde{J}_3+\delta \widetilde{J}_3 & 0 
    \end{matrix}\right),
\eeq
where
\beq
\delta\Omega = \dfrac{\widetilde{J}_2^2(\alpha_a+\alpha_b+2\Delta)}{\widetilde{J}_3^2-(\alpha_a+\alpha_b+\Delta)(\alpha_a+\alpha_b+2\Delta)},
\eeq
\beq
\delta\widetilde{J}_3 =-\frac{1}{2} \dfrac{\widetilde{J}_2^2\widetilde{J}_3}{\widetilde{J}_3^2-(\alpha_a+\alpha_b+\Delta)(\alpha_a+\alpha_b+2\Delta)},
\eeq
and generalizes \eqref{H3eff} in the presence of a detuning $\Delta$.

\bibliographystyle{apsrev4-2}
\bibliography{gmon}
\end{document}